\pdfoutput=1
\documentclass[10 pt]{article}

\usepackage{mathrsfs}
\usepackage{amssymb}
\usepackage{amsmath}
\usepackage{cite}
\usepackage{graphicx}
\usepackage[caption=false]{subfig}
\usepackage{color}
\usepackage{float}
\usepackage{multicol}
\usepackage{multirow}
\usepackage{soul}
\usepackage{bm}
\usepackage{enumerate}
\usepackage[margin=1.25in]{geometry}

\newtheorem{theorem}{Theorem}
\newtheorem{lemma}{Lemma}

\newtheorem{corollary}{Corollary}

\usepackage[table]{xcolor}
\definecolor{mygray}{gray}{.9}

\newcommand{\diag}{{\rm diag\;}}
\newcommand{\col}{{\rm col\;}}

\def\matt#1{\begin{matrix}#1\end{matrix}}
\def\qed{ \rule{.1in}{.1in}}

\begin{document}

	\title{A Distributed Algorithm for Least Square Solutions of Linear Equations}
	\author{Xuan Wang, Jingqiu Zhou, Shaoshuai Mou and Martin J. Corless
		\thanks{This work was supported by a funding from Northrop Grumman Corporation. X. Wang, J. Zhou,  S. Mou and M. J. Corless are with the School of Aeronautics and Astronautics, Purdue University, West Lafayette, IN 47906 USA (e-mail: wang3156@purdue.edu, zhou745@purdue.edu, mous@purdue.edu, corless@purdue.edu). Corresponding Author: Shaoshuai Mou.}
	}		
	\maketitle

	\begin{abstract}
		A distributed discrete-time algorithm   is proposed for multi-agent networks to achieve a common least squares solution of a group of linear equations, in which each agent only knows some of the   equations and is only able to receive information from its nearby neighbors. For  fixed, connected, and undirected networks, the proposed discrete-time algorithm 
        results in each agents solution estimate
        to converging exponentially fast 
          to the same least squares solution. Moreover, the convergence does not require careful choices of time-varying small step sizes. 
		
	\end{abstract}

	\thispagestyle{empty}
	\pagestyle{empty}

	\section{Introduction}
	{A} significant amount of effort in the control community has recently been given to distributed algorithms for solving  a set of linear equations over multi-agent networks, in which each agent only knows some of the equations and controls a state vector that can be looked upon as an estimate to the solution of the overall linear equations \cite{SJA15TAC,BSUA16NACO,JC16CNS,JN16ACC}. The key idea of these distributed algorithms is a so-called ``agreement principle" \cite{SA13ECC}, in which each agent limits the update of its state to satisfy its own equation while trying to reach a consensus with its nearby neighbors' states. 
    Different from the well-studied consensus problem \cite{AJA03TAC,ZBM07JCO,MAB08TAC,ZSBA2014CDC,MM15ECC,MAB08SIAM,XY08IFAC,KH11TAC,KH12Auto,MFMR17Auto,XJMZT17TAC,YJL06Auto}, which aims to drive all agents' states to be the same, the agreement principle allows agents to cooperatively reach the same solution to the overall set of linear equation as long   solutions exist. Numerous extensions along this direction include achieving solutions with the minimum Euclidean norm \cite{XSD17TIE,PWZ16CDC}, elimination of the initialization step \cite{LDA16ACC}, and reduction of state vector dimension by utilizing the sparsity of the linear equations \cite{SZLDA16SCL}. 
	
	Linear equations arising from many engineering problems are, however, overdetermined, for which all the above distributed algorithms based on the agreement principle are not directly applicable. For example, in distributed parameter estimation \cite{SJK12TIT}, observations are subject to measurement noise that leads to no solution of the resulting equations; in power networks, the mode estimation of voltage oscillations asks for the least squares solution of linear equations resulted from the output of phasor measurement units \cite{SJA2015TSG,BFC15CNS}; a distributed least squares solver can also be applied to the position determination of multi-agent formation control \cite{JC16Auto,ZMB04TAC,ZUB15CDC}, state estimation in signal processing \cite{TK08TSP,CA08TSP,GIG09TSP} and real-time data fitting of financial models \cite{FE01SFS}. One idea for dealing with the case of overdetermined linear equations is briefly discussed in \cite{SJA15TAC}, which however does not scale well with the network size. 
A common approach to achieve least squares solution is to reformulate it as a distributed optimization problem. In order to find the optimal solution in the  sense of the total network, classical methods employ a centralized agent (coordinator) to collect the information in the network or assign computation tasks to  other agents \cite{Z05TIT,TK08TSP}. Such a structure, however, puts too much load on the central agent and has a  strict requirement on the network topology.
Compared with this, consensus based algorithms can solve distributed optimization problem with no requirement on 
a central agent \cite{AA09TAC,JAM12TAC,DJJ14TAC,TAA14TAC,AAP10TAC,ASJH16arXiv}. For example, the methods based on the projection-consensus flow proposed in \cite{GBU17TAC,CA08TSP,GIG09TSP}, are able to drive agents' states to a neighborhood of the least square solution by introducing a sufficiently large gain. As an improvement of these methods, the exact least squares solution can be obtained by introducing a decaying weight to the local gradient\cite{ASJH16arXiv,AAP10TAC}, but at the cost of losing fixed exponential convergence rate.
Many other algorithms, like \cite{AAP10TAC,JAM12TAC,DJJ14TAC,TAA14TAC}, have good results on  both exact solution and convergence rate but require all agents to share a common, time-varying small step size that has to be carefully chosen for convergence. A similar requirement of sufficiently small step sizes has to be made when adapting classical continuous algorithms in \cite{JN12ACC,JN11ECC,BJ14TAC}  to achieve the least squares solution.
	
	The major contribution of this paper comes from devising a discrete-time algorithm, which is distributed; achieves exact least square solutions; converges exponentially fast for fixed undirected connected networks; and does not involve any small or time-varying step sizes for convergence. These attributes differentiate the proposed algorithm from those in existence for achieving a least squares solution.  The remainder of this paper is organized as follows. In Section II, we formulate the problem of obtaining aleast squares solution in a distributed manner. Then, a discrete-time distributed algorithm is proposed in Section III and  Section IV contains the main theorem which   claims exponential convergence of the proposed algorithm
    to a least squares solution. A proof of the main theorem is
    contained in Section V, which is followed by numerical simulations in Section VI and   conclusions in Section VII. Proofs of a lemma and a corollary are given in the Appendix.
	
	\smallskip
	\noindent{ \emph{Notation}:} 
    The vector ${\bf 1}_r$ is the vector in $\mathbb{R}^r$ with all its
    components equal to 1  and $I$ denotes an identity matrix. 
   The transpose and kernel of any matrix
   $M$ is denoted by $M'$, respectively.  
   By $M> 0$ and $M\geq 0$ it is meant that the symmetric matrix $M$ is positive definite and positive semi-definite, respectively.
  Finally,  
  $$\col\{A_1,A_2,\cdots,A_r\} =
    \left[\begin{array}{cccc}
    A_1'&A_2'& \cdots& A_r'\end{array}\right]'
    $$
	   and $\diag\{A_1,A_2,\cdots,A_r\}$ is  a block diagonal matrix with the $i$th diagonal block equal to $A_i$, $i=1,2,\cdots,r$. 

	\section{Problem Formulation}
	Consider a network of $m$ agents, $i=1,2,...,m$, in which each agent can  communicate with  certain other agents called its \emph{neighbors}. 
    Suppose that each agent wishes to solve   the following \emph{least squares optimization problem} 
	\begin{equation} \label{eq_orig} \min _{x\in \mathbb{R}^n} \frac{1}{2}\sum\limits_{i = 1}^m |A_ix-b_i|_2^2
	\end{equation} 
    for  the \emph{least squares solution  $x^*$},
	where $|\cdot|_2$ denotes the 2-norm,
    but   {\em each agent $i$ only knows matrices $A_i\in \mathbb{R}^{n_i\times n}$ and $b_i\in \mathbb{R}^{n_i}$.}

	Suppose that at each time
    $t=0,1, \dots$,
    each agent $i$ controls a state vector $x_i(t)\in \mathbb{R}^n$, which can be viewed  as agent $i$'s estimate of $x^*$. The \textbf{problem} of interest in this paper is to develop a local rule for each agent $i$ to iteratively update its state vector $x_i(t)$ by only using its neighbors' states such that all $x_i(t)$ converge exponentially fast to a least squares solution $x^*$.
    
      Note that, if  $x^*$ is a least squares solution then,
        all least squares solutions are given by
    \begin{equation}
    x^* + r
    \end{equation}
   with $A_i r = 0$ for $i=1, \dots, m$, that  is,
    \begin{equation}
    Ar = 0
    \end{equation}
    where
    \begin{align}
	A&=\col \{A_1,A_2,\cdots\!,A_m\}
	\end{align}%
     Thus the problem (\ref{eq_orig}) has a unique solution if and only if
    $\ker(A) =0$. 
   
   To proceed, we let  $\mathcal{N}_i$ denote the set of agent $i$'s neighbors. 
	We assume that each agent is a neighbor of itself, that is, $i\in \mathcal{N}_i$. Neighbor relations can be described by an undirected graph $\mathbb{G}$ with self-arcs such that there is an undirected edge connecting two different nodes $i$ and $j$ if and only if $i$ and $j$ are neighbors. In this paper we only consider the case in which $\mathbb{G}$ is connected and fixed.

	\section{A Distributed Discrete-Time Update}
	In this section we  present a distributed and discrete update algorithm for each agent to asymptotically achieve the same least squares solution $x^*$. 
	We note that the problem \eqref{eq_orig} is equivalent to
	the following constrained optimization problem:
	\begin{equation} \label{Objective}
	\begin{matrix}
	{\rm minimize} &\quad \displaystyle \frac{1}{2}\sum\limits_{i = 1}^m {| {{A_i}x_i - {b_i}} |_2^2} \\ 
	{{\text{subject}}\;{\text{to}}}&x_1=x_2\cdots=x_m
	\end{matrix}
	\end{equation}
	
	To obtain an update algorithm,
	let $W$ be a symmetric weighting matrix associated with the undirected graph $\mathbb{G}$ such that its $ij$-th and $ji$-th entries, $w_{ij}$ and $w_{ji}$,  are   positive    if and only if there is an undirected edge between $i$ and $j$ in $\mathbb{G}$, and are zero, otherwise. 
	Since each agent is a neighbor of itself  one has $w_{ii}>0$. Let $D$ denote the diagonal matrix whose  $i$th diagonal entry, denoted by $d_i$,   is the $i$-th row sum of $W$, that is, \begin{equation}
	d_i=\sum_{j=1}^{m}w_{ij}= \sum_{j\in \mathcal{N}_i }w_{ij}. \end{equation}
	Now introduce the Laplacian $L$  matrix associated with
	the weighted graph:
	\begin{equation}
	L =D -W
	\end{equation}
	let $\bar{L} =L  \otimes I_{n}$ where $\otimes$ denotes the Kronecker product and let \begin{equation}
	\bm{x}=\col\{x_1,x_2,\cdots,x_m\}
	\end{equation}
	be the column consisting of  all the state vectors. 
	Since $\mathbb{G}$ is connected,  a vector is the kernel of $L $ if and only if it is a scalar multiple of
    ${\bf 1}_m$  \cite{chung97GT}.
	Using this property,  one obtains that  the constraint in \eqref{Objective}
	is equivalent to ${\bar{L}\bm{x}}=0$.
	Thus   problem (\ref{Objective}) and, hence, the original problem, is equivalent to   the following problem:
	\begin{align} \label{Opt_Global}
	\begin{matrix}
	{{\text{minimize}}}\;\;&\displaystyle  \frac{1}{2}{{| {\bar A\bm{x} - b} |}_2^2} \\ 
	{{\text{subject}}\;{\text{to}}}\;\;&{{{\bar L}}\bm{x} = \bm{0}} 
	\end{matrix}
	\end{align}
	where
	\begin{align} \bar{A} &=\diag \{A_1,A_2,\cdots,A_m\} \label{eq_bA}\\
    b&=\col \{b_1,b_2,\cdots\!,b_m\}
    \label{eqbDEfn}
	\end{align} 
  Note that  ${\bm{x}^*}$ solves \eqref{Opt_Global}
   if and only if
    \begin{equation}
    {\bm{x}^*}=\bm 1_m \otimes x^*
    \end{equation}
    where $x^*$ is a least squares solution to the original problem.

	The linear constraint quadratic optimization problem (\ref{Opt_Global}) is analytically solvable by Lagrange Method\cite{B1999NL,BJ14TAC}. That is, define
	\begin{equation}
	G(\bm{x},\bm{z})=\frac{\bar{c}}{2}\left(\bar{A}\bm{x}-b\right)'\left(\bar{A}\bm{x}-b\right)+\bm{z}'\bar{L} \bm{x}
	\end{equation} \normalsize where $\bm{z}=\col\{ z_1 , z_2 , \cdots, z_m\}\in \mathbb{R}^{nm}$ is the so-called Lagrange multiplier and $\bar{c}> 0$ is an arbitrary positive constant introduced for the purpose of adjusting the weights between the two terms summed in $G(\bm{x},\bm{z})$. Note that the Hessian matrix of the objective function is $\bar{A}'\bar{A}\ge0$.
    Then,  $\bm x^*$  solves problem (\ref{Opt_Global})   if and only if there exists $\bm{z}^*$ such that ${\nabla _{\bm{x},\bm{z}}}G\left( {\bm{x}^*,\bm{z}^*} \right) = 0$. Then the problem of achieving a least square solution $x^*$ to  \eqref{eq_orig}
	is equivalent to   finding $\bm x^*$ and $\bm z^*$ such that 
	\begin{align}  \label{Eq_G1}
	\bar{c}(\bar{A}'\bar{A}\bm{x}^*-\bar{A}'b)+\bar{L} '\bm{z}^*=0 \\
	\bar{L} \bm{x}^*=0 \label{Eq_G2}
	\end{align}
	Since $W$ is symmetric, $L=L'$; hence
	\eqref{Eq_G1}-\eqref{Eq_G2} are equivalent to
	\begin{eqnarray}
	\bar{c}(A_i'A_ix_i^*-A_i'b_i) + \sum\limits_{j \in \mathcal{N}_i}w_{ij} {\left({z_i^*}-{z_j^*}\right)}&=&0 \label{Eq_L1}\\
	\sum\limits_{j \in \mathcal{N}_i}w_{ij} {\left(x_i^*-x_j^*\right)}&=&0 \label{Eq_L2}
	\end{eqnarray}
	for $i=1,2,\cdots,m$.
	
	By introducing  an additional state vector $z_i(t)\in \mathbb{R}^n$ for  each agent $i$, one could achieve a distributed solution to (\ref{Eq_L1}) and (\ref{Eq_L2}) by the saddle-point dynamics proposed in \cite{JN10AAC,BJ14TAC}. Discretization of such a continuous update usually requires a careful choice of sufficiently small step size to guarantee convergence. To eliminate such a requirement, we propose a new discrete-time update as follows:
	\begin{align}
	{x_i}\left( {t + 1} \right) =   {x_i}\left( t \right)  
    &- \bar{c} \displaystyle {\kappa}_i\left[ {A_i'{A_i}{x_i}\left( {t + 1} \right) - A_i'{b_i}} \right] \nonumber\\ 
	&-{\kappa}_i\displaystyle
	\sum\limits_{j \in {\mathcal{N}_i}} {w_{ij}
		\left[  {{z_i}\left( {t + 1} \right)  - {{{z_j}\left( t \right)}}} \right]} \nonumber\\ 
	& -   c{\kappa}_i\displaystyle \sum\limits_{j \in {\mathcal{N}_i}} {w_{ij}\left[ {{{{x_i}\left( {t + 1} \right)}} - {{{x_j}\left( t \right)}}} \right]} 
	\label{Dfomula1}
	\\   
	{z_i}\left( {t + 1} \right) = {z_i}\left( t \right) 
    &+{\kappa}_i\displaystyle \sum\limits_{j \in {\mathcal{N}_i}} {w_{ij}\left[ {{x_i}\left( {t + 1} \right) - {x_j}\left( t \right)} \right]}  \label{Dfomula2} 
	\end{align}	 
	Here $c\ge 0$ is arbitrary non-negative constant and $\kappa_i>0$, $i=1,\cdots,m$, are parameters to be chosen. As will be shown later, a simple and distributed way of choosing $c,\bar{c}, \kappa_i$ for each agent is $$c\ge 0, \quad \bar{c}>0 ,\quad \kappa_i=\frac{1}{d_i}$$ Under this choice the updates (\ref{Dfomula1})-(\ref{Dfomula2}) will be totally distributed without any designed parameters and the effectiveness for driving all $x_i(t)$ to a least square solution will be shown later in next section.  
	
	The updates (\ref{Dfomula1})-(\ref{Dfomula2}) result from a mixed use of each agent's upcoming states $x_i(t+1),z_i(t+1)$ and its neighbors current states $x_j(t),z_j(t),j\in \mathcal{N}_i$. This enables us to derive from (\ref{Dfomula1}) and (\ref{Dfomula2})  the following update without introducing any step size:
	{\small
		\begin{align}
		\label{Dupdate}
		\left[ 
		\begin{matrix}
		x_i(t + 1) \\ 
		z_i(t + 1) 
		\end{matrix} 
		\right] 
		= 
		E_i 
		\left[
		\begin{matrix}
		x_i(t)
		+  \kappa_i \sum \limits_{j \in \mathcal N_i} w_{ij}
		[ cx_j( t ) +z_j( t)]
		+ \bar{c}\kappa_i A'_{i}{b_i}  \\ 
		-  
		\kappa_i \sum\limits_{j \in \mathcal N_i} w_{ij} x_j(t) 
		+z_i ( t)
		\end{matrix} 
		\right]
		\end{align}
	}
	where
	\begin{align*}
	E_i=  \left[ \begin{matrix}
	I_{n} +   \bar{c}\kappa_i{A'_{i}}{A_i} +
	{c\kappa_i d_i  }I_{n} &
	{\kappa}_i d_i I_{n} \\ 
	- {\kappa}_i d_i I_{n} &I_{n} 
	\end{matrix} \right]^{ - 1}  
	\end{align*}
	Note right away that the update (\ref{Dupdate}) is distributed since each agent $i$ only uses $A_i,b_i$ and states of its neighbors and itself; it requires each agent to control a state vector in $\mathbb{R}^{2n}$ whose size is independent of the underlying network, and does not involve any step size. Exponential convergence under the proposed update will expounded on in next section.  

	

	\section{Main Result}
	To present the main result of the paper, Theorem \ref{T_Convergence},
	let
	$$
	\bar{W} ={W} \otimes I_{n}\,, \quad \bar{D} ={D} \otimes I_{n}\,, \quad\bar{\mathcal{K}}= \mathcal{K} \otimes I_{n} $$ where $\mathcal{K}\in\mathbb{R}^{m\times m}$ is the diagonal matrix whose $ii$ entry is $\kappa_i$.
	
	\medskip
	\begin{theorem} \label{T_Convergence}
		Suppose $\mathbb{G}$ is undirected and connected, $W$ is symmetric, $\bar{c}, \kappa_1, \dots, \kappa_m >0$,
        $c\ge0$ 
		\begin{align} \label{Condition_1}
		{D} {\mathcal{K}}{D} -{W} {\mathcal{K}}{W} \ge0 
		\end{align} 
		and  $c>0$ if there exists a non-zero vector $\bm{u}$  such that
		\begin{align} 
		\bar A\bm{u}= & 0  \label{C_lei0}\\ \label{C_lei01}
		(\bar{D} \bar{\mathcal{K}}\bar{D} -\bar{W} \bar{\mathcal{K}}\bar{W} )\bm{u}= & 0\\
		\bar{L}\bm{u} \neq& 0. \label{c_neq}
		\end{align}
		Then the proposed update (\ref{Dupdate}) results in all $x_i(t)$ converging exponentially fast to the same least squares solution to $Ax=b$.
	\end{theorem}
	\medskip
	
	By Theorem \ref{T_Convergence}, convergence of the proposed update depends on choosing   parameters $\kappa_i$ to satisfy \eqref{Condition_1}.  This can be  achieved in a simple and \textbf{distributed} way as illustrated by   the following corollary. 
    
    \medskip
	\begin{corollary} \label{L_Conditions}
		If  $c\ge0$, $\bar{c}>0$   and $\kappa_i= {1}/{d_i}$ for $i=1,2,...,m$ then the proposed update (\ref{Dupdate}) results in all $x_i(t)$ convergerging exponentially fast to the same least squares solution to $Ax=b$.
	\end{corollary}
   \medskip 
    A proof of   Corollary {\ref{L_Conditions}} is given   in the Appendix.
	
	\medskip

	\section{Proof of Theorem \ref{T_Convergence}}

	In order to prove our main result, we first  re-write   the update equations (\ref{Dfomula1})-(\ref{Dfomula2}) in vector form: 
	\begin{align}\label{GDfomula}
	\bm{x}\left( {t + 1} \right)  = \bm{x}(t)  & -  \bar{c}\bar{\mathcal{K}}\left[ {{\bar{A}'}\bar{A}\bm{x}\left( {t + 1} \right) - {\bar{A}'}b} \right] \nonumber\\
	&-  \bar{\mathcal{K}}\left[ {\bar{D} \bm{z}\left( {t + 1} \right) - \bar{W} \bm{z}\left( t \right)} \right] \nonumber\\ 
	&-  c\bar{\mathcal{K}}\left[ {\bar{D} \bm{x}\left( {t + 1} \right) - \bar{W} \bm{x}\left( t \right)} \right]\\ 
	\bm{z}\left( {t + 1} \right) = \bm{z}\left( t \right)& + \bar{\mathcal{K}}\left[ {\bar{D} \bm{x}\left( {t + 1} \right) - \bar{W} \bm{x}\left( t \right)} \right]  
    \label{GDfomula2}
	\end{align}
	where 
	\begin{align}
	\bm{x}&=\col\{x_1,\cdots,x_m\}\\
	\bm{z}&=\col\{z_1, \cdots,z_m\}
	\end{align}
    Since $\bar{\mathcal{K}}$ is invertible and $\bar{c} >0$, all
    equilibrium states $\col\{ \bm{x^e}, \bm{z^e}\}$ of \eqref{GDfomula}-\eqref{GDfomula2}
    are given by
   $ \bar{c}[\bar{A}'\bar{A}\bm{x^e} - \bar{A}'b]
    +\bar{L}\bm{z}^e + c\bar{L}\bm{x}^e =0$ and 
    $\bar{L}\bm{x}^e =0$ 
    where \begin{equation}
    \bar{L} = \bar{D} -\bar{W} = L\otimes I
    \end{equation}
    Thus, the equilibrium states are given by
    \begin{align}
     \bar{c}(\bar{A}'\bar{A}\bm{x^e} - \bar{A}'b )
    +\bar{L}\bm{z}^e &=0  \\  
     \bar{L}\bm{x}^e &=0
    \end{align}
   Clearly   the set an equilibrium states  to
    \eqref{GDfomula}-\eqref{GDfomula2} is the same as the set of solutions to 
      \eqref{Eq_G1}-\eqref{Eq_G2}.
      So, to prove Theorem \ref{T_Convergence} we just need to show
      that every solution to \eqref{GDfomula}-\eqref{GDfomula2} converges to an equilibrium state.
     To achieve this
we re-write \eqref{GDfomula}-\eqref{GDfomula2}   compactly   as
\begin{equation}
\label{eq:ysys}
 {\bm y}(t+1) =Q\bm{y} (t)+\bm{b}
\end{equation}
where ${\bm y}=\col\{\bm{x}, \bm{z}\}$
	and
	\begin{align} \label{Ex_MO}
	Q\!=\!& \left[ {\begin{matrix}
		{I_{mn}+  \bar{c}\bar{\mathcal{K}}\bar A'\bar A}\!+\!{ c\bar{\mathcal{K}}\bar{D} } &\!\!\bar{\mathcal{K}}\bar{D}  \\ 
		{ - \bar{\mathcal{K}}\bar{D} }&\!\!I_{mn} 
		\end{matrix}} \right]^{-1}\left[ {\begin{matrix}
		I_{mn}\!\!+\!c\bar{\mathcal{K}}\bar{W} & \!\!\bar{\mathcal{K}}\bar{W}  \\ 
		{ - \bar{\mathcal{K}}\bar{W} }& \!\!I_{mn}
		\end{matrix}} \right]\\
	\bm{b}\!=\!& \left[ {\begin{matrix}
		{I_{mn}+  \bar{c}\bar{\mathcal{K}}\bar A'\bar A}+{ c\bar{\mathcal{K}}\bar{D} } &\bar{\mathcal{K}}\bar{D}  \\ 
		{ - \bar{\mathcal{K}}\bar{D} }&I_{mn} 
		\end{matrix}} \right]^{-1} \left[ {\begin{matrix}
		\bar{c}\bar{\mathcal{K}}\bar A'b \\ 
		0
		\end{matrix}} \right]
	\end{align} 

 The equilibrium states $\bm{y^e}$ of \eqref{eq:ysys} are given by
$
 (I-Q)\bm{y^e} = \bm{b}
$.
Let $ \bm{y^*} $  be any equilibrium state of  \eqref{eq:ysys}.
Then all equilibrium states $\bm{y^e}$  of \eqref{eq:ysys} are given by
$
\bm{y^e} =  \bm{y^*} + \bm{v}
$
where $\bm{v} =0$  or $\bm{v}$ an eigenvector  of $Q$ corresponding to  eigenvalue one.
The evolution of 
$
\bm{e}= 	\bm{y} - \bm{y^*}
$
is governed by
\begin{equation}
\label{eq:esys}
 {\bm e}(t+1) = Q \bm{e}(t)
\end{equation}
So, to prove Theorem \ref{T_Convergence} we now  just have to show that every solution of 
\eqref{eq:esys}
 exponentially converges to zero or to  an eigenvector  of $Q$ corresponding to  eigenvalue one. To achieve this we need 
 the following lemma whose proof is in the Appendix


  \medskip
	\begin{lemma} \label{lem:eigQ} 
		Suppose   (\ref{Condition_1}) holds. Then $Q$ has the following properties.
        \begin{itemize}
        \item[(a)]
        Every eigenvalue of $Q$ has magnitude less than or equal to one and $-1$ is not an eigenvalue of $Q$.
        \item[(b)]
          If  $Q$ has a  complex eigenvalue 
			of magnitude one 
			  then  $c=0$ and there is a non-zero vector $\bm{u}$ which satisfies   (\ref{C_lei0}) - (\ref{c_neq}) in Theorem \ref{T_Convergence}. 
              \item[(c)]
              One  is  an eigenvalue of $Q$  and its algebraic multiplicity is equal to its geometric multiplicity.
              A non-zero  vector $\col\{\bm{u},\bar{\bm{u}}\}$ is a   eigenvector corresponding to one
			if and only if 
            \begin{equation} \label{cond:eig=1}
            \bar A\bm{u}=0,\qquad 
            \bar{L}\bm{u} = 0,\qquad 
            \bar{L}{\bar{\bm{u}}}=0
            \end{equation}
        \end{itemize}
	\end{lemma}
	%


	
	\medskip

As a consequence of the hypotheses of Theorem \ref{T_Convergence}, Lemma \ref{lem:eigQ}   tells us that every eigenvalue of
$Q$ has magnitude less than or equal to one.
Also, one is the only eigenvalue of magnitude one and its algebraic and geometric multiplicities are equal.
Hence, there exists a non-singular matrix $T$ such that
\begin{align} \label{trans}
	Q= T \left[\matt{I & 0\\ 0 & R}\right]  T ^{-1}
	\end{align} 
	and all the eigenvalues of $R$ have magnitude strictly less than one.
	Every solution of \eqref{eq:esys} satisfies
	$\bm{e}(t) = Q^t\bm{e}(0)$.
	Since 
	\[
	Q^t =  T \left[\matt{I & 0\\ 0 & R^t}\right]  T ^{-1}
	\]
	and  all the eigenvalues of $R$ have magnitude strictly less than one,  it follows that $\bm{e}(t)$ exponentially converges to  
	\[
	\bm{v} =   T \left[\matt{I & 0\\ 0 & 0}\right]  T ^{-1}\bm{e}(0)
	\]
	Note that
	\begin{align*}
	Q\bm{v} &= T \left[\matt{I & 0\\ 0 & R}\right]  T ^{-1}T \left[\matt{I & 0\\ 0 & 0}\right]  T ^{-1}\bm{e}(0)\\
    &= T \left[\matt{I & 0\\ 0 & 0}\right]  T ^{-1}\bm{e}(0) =\bm{v}
	\end{align*}
	that is,  $\bm{v}=0$ or $\bm{v}$ is an eigenvector  of $Q$ corresponding to  eigenvalue one.
	Hence every solution of 
\eqref{eq:esys}
exponentially  converges to zero or to  an eigenvector  of $Q$ corresponding to  eigenvalue one.\qed

	\section{Example} \label{Sec_Val}
\begin{figure}[th]
		\centering 
		\includegraphics[width=7 cm]{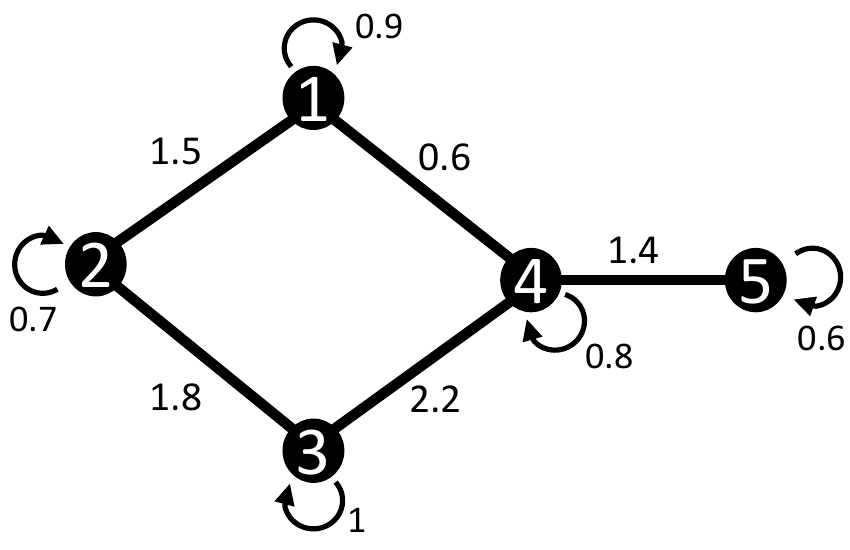}
		\caption{Five-node connected network $\mathbb{G}$}
		\label{Fig1}
	\end{figure}

	Numerical simulations will be performed with the 5-node network in Fig. \ref{Fig1} to illustrate  Theorem \ref{T_Convergence}. 
    From Fig. \ref{Fig1} we obtain that
    \begin{align*} 
	W=&
	\begin{bmatrix}
	0.9 & 1.5 &   0 & 0.6 &   0\\
	1.5 & 0.7 & 1.8 &   0 &   0\\
	0 & 1.8 &   1 & 2.2 &   0\\
	0.6 &   0 & 2.2 & 0.8 & 1.4\\
	0   &   0 &   0 & 1.4 & 0.6\\
	\end{bmatrix} 
    \end{align*}
    Hence
   \[
	D =
	\diag\left\{\begin{matrix}3, &4, & 5, & 5, & 2 \end{matrix}\right\} 
    \]
and based  on the distributed way of choosing weights in  Corollary \ref{L_Conditions},   $$ \mathcal{K}=
	\diag\left\{\begin{matrix}\displaystyle \frac{1}{3}, &\displaystyle \frac{1}{4}, & \displaystyle \frac{1}{5}, & \displaystyle \frac{1}{5}, & \displaystyle \frac{1}{2} \end{matrix}\right\} $$
	
\begin{figure}[h]
		\vspace{-0.1cm}
		\centering 
		\includegraphics[width=8.5 cm]{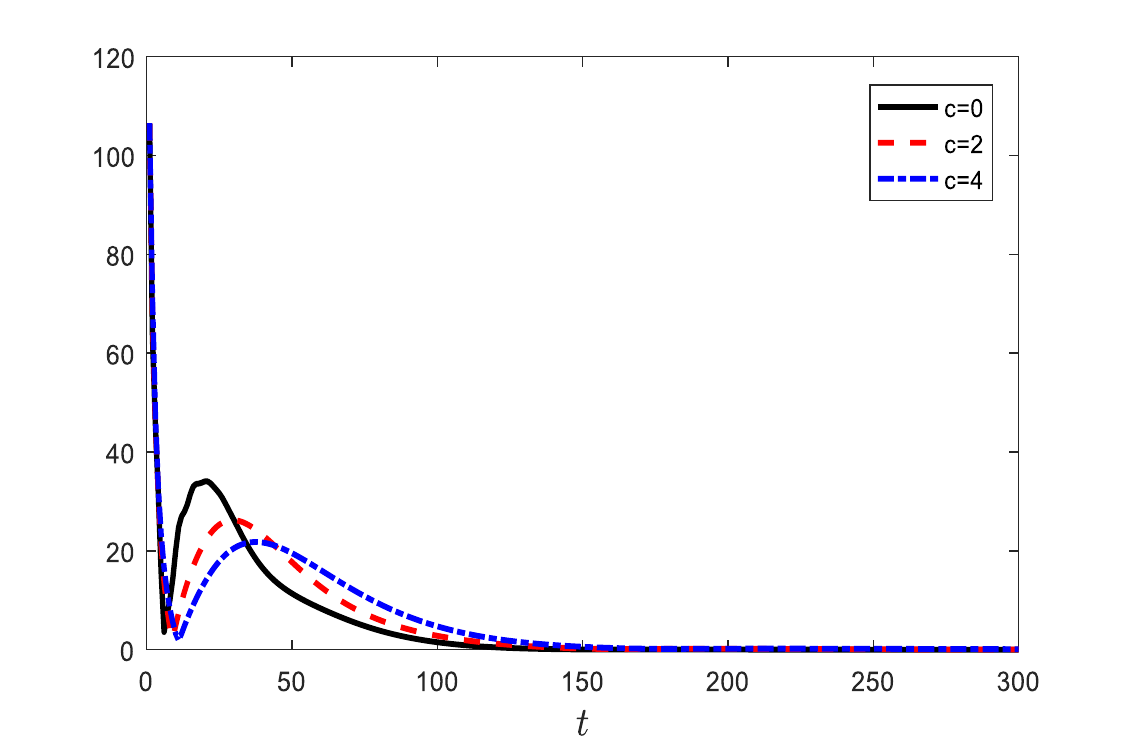}
		\caption{Example: multiple least squares solution, $\bar{c}=1$, different $c$}
		\label{Fig3}
	\end{figure}

	We consider solving  a set of linear equations that has multiple least square solutions on network $\mathbb{G}$, in which agents 1, 2, 3, 4, 5 know
	\begin{align*}
	A_1=&\left[\matt{1&2&3&4}\right],\quad b_1=10\\ 
	A_2=&\left[\matt{4&5&6&7}\right],\quad  b_2=20\\
	A_{3}=&\left[\matt{1&2&3&4}\right], \quad b_{3}=15\\
	A_{4}=&\left[\matt{5&6&3&4}\right],\quad  b_{4}=17\\
	A_{5}=&\left[\matt{4&3&2&1}\right],\quad  b_{5}=6
	\end{align*}
	\smallskip
	respectively. Since the least squares solution is not unique, to show the effectiveness of our method, we introduce $W(t)$ as
	\begin{align*} 
	W(t)\!=\!\displaystyle \frac{1}{2m}\!\!\sum_{i=1}^m\displaystyle\!|A'Ax_i(t)\!-\!A'b|_2^2 \!+\! \displaystyle \frac{1}{2m^2}\!\!\sum_{i=1}^m \!\sum_{j=1}^m \displaystyle\!|x_i(t)\!-\!x_j(t)|_2^2
	\end{align*}
	where the first term is a cost associated
    with the $x_i(t)$ not being a least squares solution of  $Ax=b$; the second term is a cost associated with
    the  $x_i(t)$  not achieving consensus to the same value.
	Numerical simulation results in Fig. \ref{Fig3} validates the exponential convergence of $W(t)$ for $\bar{c}=1$ and $c=0$, $2$, $4$, respectively.	
	\begin{figure}[h]
		\vspace{-0.1cm}
		\centering 
		\includegraphics[width=8.5 cm]{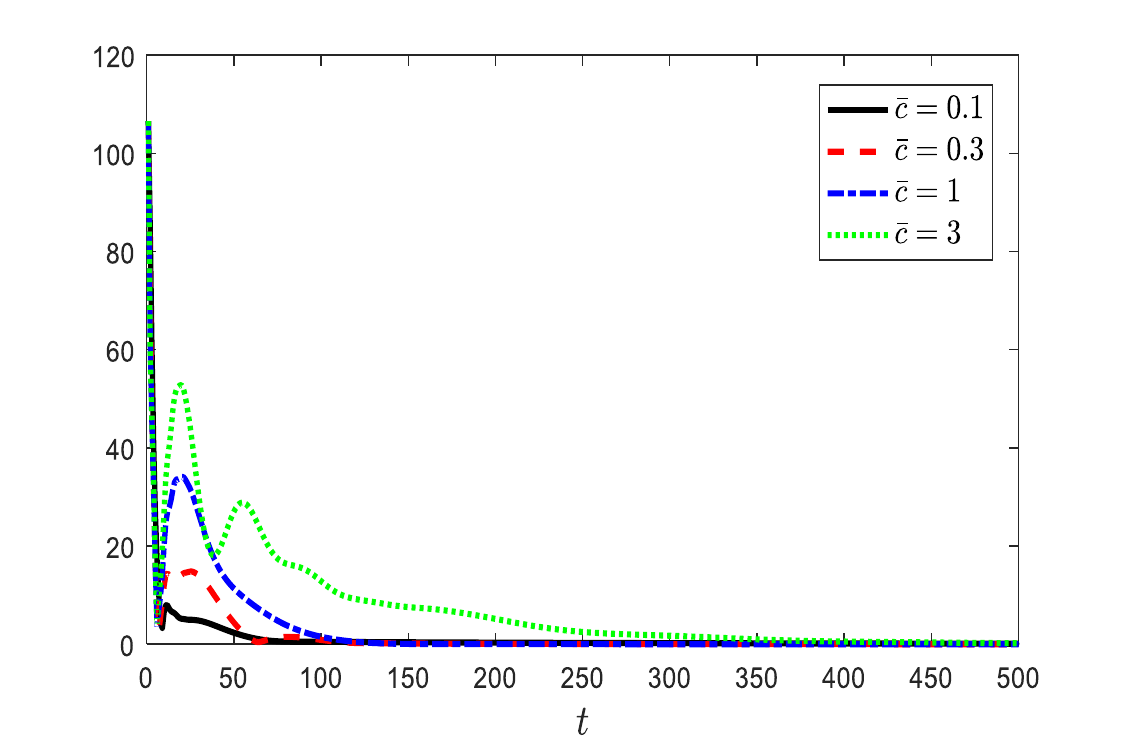}
		\caption{Example: Multiple least squares solution, $c=0$, different $\bar{c}$ }
		\label{Fig4}
	\end{figure}
	As a comparison, Fig. \ref{Fig4} validates the exponential convergence of $W(t)$ for $c=0$ and $\bar{c}=0.1$, $0.3$, $1$, $3$, respectively.
	
\begin{figure}[h]
		\vspace{-0.1cm}
		\centering 
		\includegraphics[width=8.5 cm]{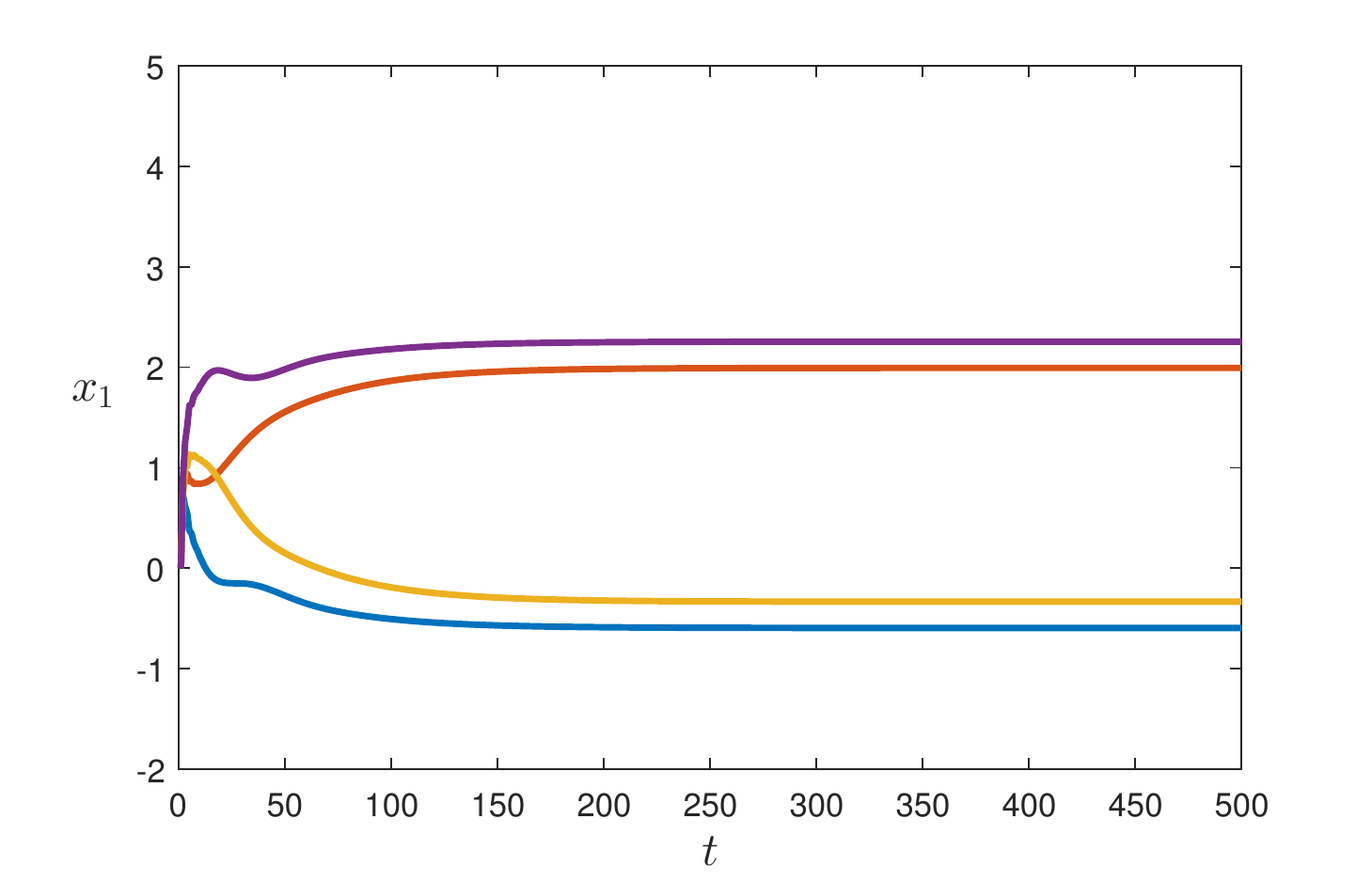}
		\caption{History of $x_1$ when $c=0$ and $\bar{c}=1$}
		\label{Figstable}
	\end{figure}

It is worth mentioning that $W(t)$ goes to $0$ doesn't mean $x_i$ converge to a constant value, to show this, we let $c=0, \bar{c}=1$ and use Fig. \ref{Figstable} to demonstrate the history of $x_1$. This, along with the consensus result in Fig. \ref{Fig3} and \ref{Fig4} validates that all $x_i$ converge to constant values.

The simulation results show that different parameters $c$ and $\bar{c}$ lead to different convergence rates, this because the eigenvalues of matrix $Q$ are different. However, finding the best parameter set $c$ and $\bar{c}$ is not straightforward because the eigenvalues of matrix $Q$ are also determined by the equation $Ax=b$ and the network $\mathbb{G}$, both information are global information that cannot be obtained by agents. 

	\section{Conclusion}
	We have proposed a discrete-time update for multi-agent networks, which enable each agent to achieve the same least square solutions exponentially fast when the network is undirected and connected. The exponential convergence does not rely on any time-varying and small step-size, which differs form the proposed updates   in existence. Future work includes proper design of parameters $c$, $\bar{c}$ and the generalization of the proposed update to networks that are directed and time-varying. 
	
	\section*{Appendix}
	\subsection*{Proof of Lemma \ref{lem:eigQ}}
    First note that due to the mixed-product property of Kronecker product,  $\bar{D} \bar{\mathcal{K}}\bar{D}=(D\mathcal{K}D)\otimes I_n$; $\bar{W} \bar{\mathcal{K}}\bar{W}=(W\mathcal{K}W)\otimes I_n$. Assumption \eqref{Condition_1} in Theorem \ref{T_Convergence} is equivalent to
$$\bar{D} \bar{\mathcal{K}}\bar{D} -\bar{W} \bar{\mathcal{K}}\bar{W} \ge0 $$
              
If suppose  $\lambda$ is  an eigenvalue of $Q$.
	Then there is a   nonzero  vector $\col\{\bm u, \bar{\bm{u}}\}$ such that
	\begin{align}
	Q\left[ \matt{\bm{u} \\ {\bar{\bm{u}}} 
	} \right] = {\lambda}\left[ \matt{\bm{u} \\ {\bar{\bm{u}}} } \right]\nonumber
	\end{align}
	which, recalling (\ref{Ex_MO}), is equivalent to 
	\begin{align*}
	\left[ {\begin{matrix}
		I_{mn}\!+\!c\bar{\mathcal{K}}\bar{W}\!\!\!\! &\bar{\mathcal{K}}\bar{W}  \\ 
		{ - \bar{\mathcal{K}}\bar{W} }\!\!\!\!&I_{mn}
		\end{matrix}} \right] \left[ \matt{\bm{u} \\ {\bar{\bm{u}}} 
	} \right]
	\!=\! {\lambda}\left[ {\begin{matrix}
		{I_{mn}\!+\!\bar{c}\bar{\mathcal{K}}\bar A'\bar A\!+\!c\bar{\mathcal{K}}\bar{D} }\!\!\!\!&\bar{\mathcal{K}}\bar{D}  \\ 
		{ - \bar{\mathcal{K}}\bar{D} }\!\!\!\!&I_{mn} 
		\end{matrix}} \right]\left[ \matt{\bm{u} \\ {\bar{\bm{u}}} 
	} \right]
	\end{align*}
	that is,
	\begin{align*}
	\left[ I_{mn}\!+\!c\bar{\mathcal{K}}\bar{W}  \!-\!  \lambda(I_{mn}+ \bar{c}\bar{\mathcal{K}}\bar A'\bar A \!+\!c\bar{\mathcal{K}}\bar{D} ) \right]\bm{u} 
	&= \!\bar{\mathcal{K}}\left( {\lambda \bar{D} - \bar{W} } \right){\bar{\bm{u}}}  \\
	\bar{\mathcal{K}}\left( {{\lambda}\bar{D}  - \bar{W} } \right)\bm{u} & = \left( {\lambda- 1} \right){\bar{\bm{u}}} 
	\end{align*}
	and, since $\bar{\mathcal{K}}$ is nonsingular, these can be written as
	\begin{align}
	\left[ {\bar{\mathcal{K}}^{-1}\!+\!c\bar{W}  \!-\!  \lambda\left( \bar{\mathcal{K}}^{-1}\!+\! \bar{c}\bar A'\bar A \!+\!c\bar{D} \right)} \right]&\bm{u}=\left( {\lambda \bar{D} - \bar{W} } \right){\bar{\bm{u}}} \label{uvsub} \\
	\left( {{\lambda}\bar{D}  - \bar{W} } \right)\bm{u} =& \left( {\lambda- 1} \right)\bar{\mathcal{K}}^{-1}{\bar{\bm{u}}} \label{uvr} 
	\end{align}
	
	In the case of $\lambda\neq 1$, \eqref{uvr} is equivalent to  
	\begin{align}
	\label{eq_uu}\bar{\bm{u}}=\frac{1}{\lambda -1} \bar{\mathcal{K}}\left( {{\lambda}\bar{L} } \right)\bm{u}
	\end{align}
	Hence $\bm{u} \neq 0$ and 
	(\ref{uvsub}) is equivalent to
	\begin{align}
	M(\lambda)\bm{u}=0 \label{eq_Mu} 
	\end{align}
	where 
	\small
	\begin{align}
	M(\lambda) &= 
	\left(  {\lambda}\bar{D}  - \bar{W}  \right) \bar{\mathcal{K}} \left(  {\lambda}\bar{D}  - \bar{W}  \right) 
	+ \lambda\left(  \lambda -\!1\right)\bar{c}\bar A'\bar A  \nonumber \\
	& +\left. {{\left( {\lambda - 1  } \right)}^2}\bar{\mathcal{K}}^{-1}+(\lambda-1)c(\lambda \bar{D} -\bar{W} \right) \label{eq_M}
	\end{align}
	\normalsize
	Note  that
	$$M(\lambda)=\lambda^2M_2+\lambda M_1+M_0$$
	where
	\begin{align*}
	M_2=&\;\bar{D} \bar{\mathcal{K}}\bar{D} +\bar{\mathcal{K}}^{-1}+c\bar{D} +\bar{c}\bar A'\bar A>0\\
	M_1=&-\bar{W} \bar{\mathcal{K}}\bar{D} -\bar{D} \bar{\mathcal{K}}\bar{W} -2\bar{\mathcal{K}}^{-1}-c\bar{D} -c\bar{W} -\bar{c}\bar A'\bar A\\
	M_0=&\;\bar{W} \bar{\mathcal{K}}\bar{W} +\bar{\mathcal{K}}^{-1}+c\bar{W} >0
	\end{align*}
	Thus, we have shown that $\lambda \neq 1$ is an eigenvalue of $Q$ if and only if there is a nonzero vector $\bm{u}$ 
	such that \eqref{eq_Mu} holds.
	
	In the case of $\lambda =1$, equations \eqref{uvsub} and \eqref{uvr} reduce to
	\begin{align*}
	-\left[ {      \bar{c}\bar A'\bar A  + c(\bar{D} -\bar{W})  } \right] \bm{u} &=\left( {  \bar{D} - \bar{W} } \right){\bar{\bm{u}}}  \\
	\left( { \bar{D}  - \bar{W} } \right)\bm{u} &=  0 
	\end{align*}
	Recall that $\bar{D}  - \bar{W}=\bar{L}$, thus $\lambda =1$ is an eigenvalue of $Q$ and 
	$\col\{\bm{u}, \bar{\bm{u}}\}$ is an eigenvector corresponding
	to $1$ if and only if $\col\{\bm{u}, \bar{\bm{u}}\}$ is nonzero and
	\begin{align}
	\bar{L}\bm{u} &=  0 \label{uvr2}\\
	\bar{L}{\bar{\bm{u}}}  &=
	-  {      \bar{c}\bar A'\bar A  }   \bm{u}
	\label{uvsub2} 
	\end{align}
	
	\subsection*{Proof of (a)}  
	
	If $|\lambda|>1$, then we have shown in the previous section that that $\lambda $ is an eigenvalue of $Q$ if and only if there is a nonzero vector $\bm{u}$ 
	such that \eqref{eq_Mu} holds.
	
	Suppose  $\lambda$ is real with $|\lambda| >1$ and
	recall expression \eqref{eq_M} for   $M(\lambda)$.
	Observe that both $\lambda(\lambda-1)$ and  $(\lambda-1)^2$ are positive; $\bar A'\bar A$ and $\left( {{\lambda}\bar{D}  - \bar{W} } \right) \bar{\mathcal{K}} \left( {{\lambda}\bar{D}  - \bar{W} } \right)$ are positive semi-definite; and $\bar{\mathcal{K}}^{-1}$ is positive definite. Furthermore, if $\lambda>1$ then, $\lambda-1$ is positive and $c(\lambda \bar{D} -\bar{W} )$ is positive semi-definite; if $\lambda<-1$ then, $\lambda-1$ is negative and $c(\lambda \bar{D} -\bar{W} )$ is negative semi-definite. Thus, we   conclude  that when $\lambda$ is real with $|\lambda| >1$, the matrix $M(\lambda)$ in (\ref{eq_M}) is   positive definite. Hence there is not a non-zero vector $\bm{u}$
	for which $M(\lambda)\bm{u} =0$ and so
	$\lambda$ is not an eigenvalue of $Q$.
	
	Suppose  that $\lambda$ is  complex.
	Left-multiplying  equation (\ref{eq_Mu}) by
	$\bm{u}'$   yields 
	$$\lambda^2c_2+\lambda c_1+c_0=0$$
	where
	\begin{align} \label{eq_c123}
	c_0=\bm{u}'M_0\bm{u},\quad c_1=\bm{u}'M_1\bm{u},\quad c_2=\bm{u}'M_2\bm{u}
	\end{align}
	Since   $ M_0, M_1, M_2$ are symmetric, $c_0, c_1$ and $c_2$ are real.
	Let $\lambda=p+q\bm{i}$, where $\bm{i}=\sqrt{-1}$     and $p$ and $q$ are real with $q\neq 0$.
	Then equating the real and imaginary parts of (\ref{eq_c123}) to zero results in
	\begin{align} \label{cpqr}
	(p^2-q^2)c_2+pc_1+c_0&=0\\
	q(2pc_2+c_1)&=0\label{cpqi}
	\end{align}
	Since $q \neq 0$,  equation (\ref{cpqi})   implies that 
	\begin{equation}
	\label{eq:pc2c1}
	2pc_2+c_1 =0
	\end{equation}
	which upon substitution into    (\ref{cpqr}) yields 
	\begin{align} \label{c1c2}
	|\lambda|^2c_2-c_0=0
	\end{align}
	Since $M_2$ is positive definite and $\bm{u}\neq0$,
	we must have $c_2>0$.
	Note that $c_2-c_0 =\bm{u}'(M_2-M_0)\bm{u}$ and
	\begin{equation} \label{eq:M2-M0}
	M_2 - M_0 = \bar{D}\bar{\mathcal{K}}\bar{D} -\bar{W} \bar{\mathcal{K}}\bar{W} +c (\bar{D} -\bar{W}) + \bar{c} \bar{A}'\bar{A}
	\end{equation}
	Since $\bar{D} -\bar{W}=\bar{L}\ge 0, \bar{A}'\bar{A} \ge 0$,
	$c, \bar{c} \ge0$ and, by assumption
	$\bar{D} \bar{\mathcal{K}}\bar{D} -\bar{W} \bar{\mathcal{K}}\bar{W}\ge 0$, one has $M_2-M_0\ge0$ and
	$c_2 \ge c_0$. Since $c_2 >0$,
	it now follows from   (\ref{c1c2}) that $|\lambda|^2 =c_0/c_2 \leq 1$.
	Hence $|\lambda| \le 1$.
	
	Recall that   $\lambda =-1$ is an eigenvalue of $Q$ if and only if there is a nonzero vector $\bm{u}$ 
	such that $M(-1)\bm{u} =0$. 
	From \eqref{eq_M} we have
	\small
	\begin{align*}
	M(-1)  &= 
	\left( \bar{D}  + \bar{W}  \right) \bar{\mathcal{K}} \left(  \bar{D}  + \bar{W}  \right) 
	+ 2 \bar{c}\bar A'\bar A   
	+  4\bar{\mathcal{K}}^{-1}+
	2c( \bar{D}+\bar{W} ) \\ 
	&>0
	\end{align*}
	\normalsize
	Since $M(-1)$   is positive definite, $M(-1)\bm{u} \neq0$
	for all non-zero $\bm u=0$.   Thus $  -1$ is not an eigenvalue of $Q$.

	\subsection*{Proof of (b)}

	Suppose that $\lambda$ is a complex eigenvalue of $Q$
	with $|\lambda| = 1$.
	Recalling the proof of (a), there must exist a 
	nonzero vector $\bm{u}$ such that 
	\eqref{eq:pc2c1} and \eqref{c1c2} hold,
	that is,
	\begin{align}
    \bm{u}'(2pM_2+M_1)\bm{u} &=0	\label{eq:2pM2M1}\\
	\bm{u}'(M_2-M_0)\bm{u} &=0		\label{eq:M2M02}
	\end{align}
	where   $\lambda=p+q\bm i$.
	Recall \eqref{eq:M2-M0}.
	Since  $c\left( \bar{D}  - \bar{W}  \right)$ and $\bar{c}\bar A'\bar A$ are positive semi-definite,  $\bar{c} >0$ and, by assumption,
	$\bar{D}\bar{\mathcal{K}}\bar{D} - \bar{W}\bar{\mathcal{K}}\bar{W}$ is positive semi-definite,
	\eqref{eq:M2M02} implies that
	\begin{align}
	(\bar{D}\bar{\mathcal{K}}\Bar{D} - \Bar{W}\bar{\mathcal{K}}\bar{W})\bm{u} =0\label{eq:DKD-WKW=0}\\
	\bar A\bm{u}=0   \label{eq:Au=0}\\
	c\left( \bar{D}  - \bar{W}  \right)\bm{u}=0
	\label{(D-W)u=0}
	\end{align}
	Equation \eqref{eq:2pM2M1}, along with  \eqref{eq:Au=0}, results in
	\begin{align} \label{fullm}
	&\bm{u}'[2p(\bar{D} \bar{\mathcal{K}}\bar{D} +\bar{\mathcal{K}}^{-1}+c\bar{D} )
	\nonumber \\  
	&-(\bar{W} \bar{\mathcal{K}}\bar{D} +\bar{D} \bar{\mathcal{K}}\bar{W} +2\bar{\mathcal{K}}^{-1}+c\bar{D} +c\bar{W} ) ]\bm{u}=0
	\end{align}
	If $c\neq 0$ then \eqref{(D-W)u=0}  implies that $\bar{D} \bm{u} = \bar{W} \bm{u}$
	and we obtain that
	\begin{align} \label{contra}
	2(p-1)\bm{u}'[ \bar{D} \bar{\mathcal{K}}\bar{D} \!+\!  \bar{\mathcal{K}}^{-1}\!+\! c\bar{D} ]\bm{u}=0
	\end{align}
    Since $1 = |\lambda|^2 = p^2+q^2=1$ and $q\neq 0$,
    we must have $p<1$ and \eqref{contra} implies that
    	\begin{align} \label{contra2}
	 \bm{u}'[ \bar{D} \bar{\mathcal{K}}\bar{D} \!+\!  \bar{\mathcal{K}}^{-1}\!+\! c\bar{D} ]\bm{u}=0
	\end{align}
    The matrix $\bar{D} \bar{\mathcal{K}}\bar{D} \!+\!  \bar{\mathcal{K}}^{-1}\!+\! c\bar{D}$ is positive definite, so,
    \eqref{contra2} yields the contradiction that
	  $\bm{u} =0$; hence $c=0$.
      
  If $\bar{L}\bm{u} =0$ then  $\bar{D} \bm{u} = \bar{W} \bm{u}$ and \eqref{contra} holds. Again we get    the contradiction that $\bm{u} =0$. Hence $\bar{L}\bm{u} \neq 0$. This along with \eqref{eq:DKD-WKW=0}, \eqref{eq:Au=0}, lead to the equations (\ref{C_lei0})-(\ref{c_neq}) in Theorem \ref{T_Convergence}. 
 \qed
	
	\medskip
	
\subsection*{Proof of (c)} 
    We have seen that one  is an eigenvalue for $Q$ and 
	$\col\{\bm{u}, \bar{\bm{u}}\}$ is a  corresponding eigenvector
	if and only if $\col\{\bm{u}, \bar{\bm{u}}\}$ is nonzero
	and satisfies \eqref{uvr2} and \eqref{uvsub2}.
	Since $\bar{L}  $ is symmetric,    \eqref{uvr2}   implies that that $\bm{u}'\bar{L} =0$. 
	Multiplying  both sides of   (\ref{uvsub2}) by 
	$\bm{u}'$, it now follows that   $-\bar{c}\bm{u}' \bar{A}'\bar A\bm{u}=0$;
	since  $\bar{c}>0$, this is equivalent to $\bar{A}\bm{u}=0$. 
	Equation \eqref{uvr2} now implies that $\bar{L}\bar{\bm{u}} = 0$.
	Thus \eqref{cond:eig=1} holds. In addition, since $\bar{L}$ is singular,  a nonzero solution $\col\{\bm{u}, \bar{\bm{u}}\}$  to \eqref{cond:eig=1} exists;
    hence one is an eigenvalue for $Q$.
	
	\smallskip
    
Now, we prove the multiplicity property of eigenvalue one by contradiction. Suppose the algebraic multiplicity of  the eigenvalue one   is not equal to its geometric multiplicity. 
    Then there exists a  non-zero vector $\col\{\bm{v},\bar{\bm{v}}\}$ \cite{Bronson91matrix} such that
	\begin{align} \label{eq_eig}
	\left(Q- I \right)\left[ \matt{
		\bm{v} \\ 
		{\bar{\bm{v}}} 
	} \right] = \left[ \matt{\bm{u} \\ 
	{\bar{\bm{u}}} 
} \right]
\end{align} 
where $\col\{\bm{u},\bar{\bm{u}}\}$
is an eigenvector corresponding to one.
It follows from   (\ref{eq_eig}) and the definition of $Q$ in (\ref{Ex_MO}) that
\begin{align*} 
(c\bar{\mathcal{K}}\bar{W}-{ c\bar{\mathcal{K}}\bar{D} } & -\bar{c}\bar{\mathcal{K}}\bar A'\bar A)  \bm{v}
- \left(\bar{\mathcal{K}}\bar{W} -\bar{\mathcal{K}}\bar{D} \right)\bar{\bm{v}} \\
&= \left(I + \bar{c}\bar{\mathcal{K}}\bar A'\bar A+c\bar{\mathcal{K}}\bar{D} \right)\bm{u}+ \bar{\mathcal{K}}\bar{D} \bar{\bm{u}}\\
(\bar{\mathcal{K}} \bar{D} -\bar{\mathcal{K}}\bar{W} )\bm{v} &=-\bar{\mathcal{K}}\bar{D} \bm{u}+ \bar{\bm{u}}
\end{align*}
Left multiplying  both equations by $\bar{\mathcal{K}}^{-1}$ 
and recalling that $\bar{L}= \bar{D} -\bar{W}$ yields
\begin{align} 
-c(\bar{L} +\bar A'\bar A )  \bm{v} +\bar{L} \bar{\bm{v}}  
 = \left(\bar{\mathcal{K}}^{-1} + \bar{c}\bar A'\bar A+c\bar{D} \right)\bm{u}+ \bar{D} \bar{\bm{u}} \label{ido1}\\
 \bar{L}\bm{v}=-\bar{D} \bm{u}+ \bar{\mathcal{K}}^{-1}\bar{\bm{u}} \label{ido2}
\end{align}
Since   $\col\{\bm{u},\bar{\bm{u}}\}$
is an eigenvector corresponding to one, it is nonzero and satisfies \eqref{cond:eig=1};
hence left multiplying  (\ref{ido1}) and (\ref{ido2})
by $\bm{u}'$ and  $\bar{\bm{u}}'$, respectively,
results in
\begin{align*} 
0 &= \bm{u}'\left(\bar{\mathcal{K}}^{-1}+c\bar{D} \right)\bm{u}+ \bm{u}'\bar{D} \bar{\bm{u}} 
\\
0 &=-\bar{\bm{u}}'\bar{D} \bm{u}+ \bar{\bm{u}}'\bar{\mathcal{K}}^{-1}\bar{\bm{u}} 
\end{align*} 
Thus
\begin{align} 
\bm{u}'\left(\bar{\mathcal{K}}^{-1}+c\bar{D} \right)\bm{u}+ \bar{\bm{u}}'\bar{\mathcal{K}}^{-1}\bar{\bm{u}}=0
\end{align} 
Since both $\left(\bar{\mathcal{K}}^{-1}+c\bar{D} \right)$ and $\bar{\mathcal{K}}^{-1}$ are positive definite,  we obtain the contradiction that $\bm{u}=\bar{\bm{u}}=0$. Hence
  the algebraic multiplicity of eigenvalue one must equal  its geometric multiplicity.  \qed

\medskip

\subsection*{Proof of Corollary \ref{L_Conditions}}
We prove this corollary by showing that the hypotheses of Theorem \ref{T_Convergence} hold.
	Since $\kappa_i={1}/{d_i}$
    we have $\bar{\mathcal{K}}=\bar{D} ^{-1}$ 
    and 
	\begin{align*}
\bar{D}  \bar{\mathcal{K}} \bar{D} -\bar{W}  \bar{\mathcal{K}} \bar{W} 
	&= \bar{D}  -\bar{W}  \bar{D} ^{-1} \bar{W} \\
	&=\bar{D} ^{\frac{1}{2}} [I-  (\bar{D} ^{-\frac{1}{2}} \bar{W}  \bar{D} ^{-\frac{1}{2}} )^{2} ] \bar{D} ^{\frac{1}{2}}
	\end{align*}
	By the Gershgorin Disk Theorem,   
    $\bar{L} = {\bar D} -{\bar W}\ge0$ \cite{chung97GT} and thus 
 $I \ge \bar{D} ^{-\frac{1}{2}} \bar{W}  \bar{D} ^{-\frac{1}{2}}$.
    Hence $I \ge (\bar{D} ^{-\frac{1}{2}} \bar{W}  \bar{D} ^{-\frac{1}{2}})^2$
   that is,
	$$I -(\bar{D} ^{-\frac{1}{2}} \bar{W}  \bar{D} ^{-\frac{1}{2}} )^{2}\ge 0$$
	Thus, $\bar{D}  \bar{\mathcal{K}} \bar{D} -\bar{W}  \bar{\mathcal{K}} \bar{W} \ge 0$, that is, (\ref{Condition_1}) holds. 
	
	\medskip
	
	We now show that equations (\ref{C_lei0})-(\ref{c_neq}) do not have a   solution. If (\ref{C_lei01})
    holds then,
	\begin{align*}
	\left(\bar{D}-\bar{W} \bar{D} ^{-1}\bar{W}\right) \bm{u}= 0
	\end{align*}
that is 
\begin{align} \label{expu1}
 	\bar{D} ^{\frac{1}{2}}\!\left(I \!+\!\bar{D} ^{-\frac{1}{2}}\bar{W}\bar{D} ^{-\frac{1}{2}}\right)\!\left(I\!-\!\bar{D} ^{-\frac{1}{2}}\bar{W}\bar{D} ^{-\frac{1}{2}}\right)  \bar{D} ^{\frac{1}{2}}\bm{u}=0
    \end{align}
	Recall that $\mathbb{G}$ has self-arcs, that is, $\bar{D} +\bar{W} >0$ \cite{chung97GT};
    hence   $I_{mn}\!+\!\bar{D} ^{-\frac{1}{2}}\bar{W}\bar{D} ^{-\frac{1}{2}}>0$ is nonsingular. Thus, equation \eqref{expu1} leads to 
 \begin{align} \label{expu2}
 	\left(I_{mn}-\bar{D} ^{-\frac{1}{2}}\bar{W}\bar{D} ^{-\frac{1}{2}}\right)  \bar{D} ^{\frac{1}{2}}\bm{u}=0
    \end{align}   
which, upon    left-multiplying by  $\bar{D} ^{\frac{1}{2}}$
yields
 \begin{align} \label{expu3}
 	\left(\bar{D}-\bar{W}\right)\bm{u}=0
    \end{align} 
    that is
   \eqref{c_neq} is not satisfied.
   Hence there does not exist a vector $\bm{u}$ satisfying (\ref{C_lei0}) - (\ref{c_neq}). 
   Application of Theorem \ref{T_Convergence} now yields exponential convergence for any $c\ge0$ and $\bar{c}>0$. \qed


\bibliographystyle{Bibliography/IEEEtran}
\bibliography{Bibliography/IEEEabrv,Bibliography/Shaoshuai,Bibliography/LinearEquations,Bibliography/Friends,Bibliography/IndusApps,Bibliography/LeastSquare,Bibliography/Xuanwang}

\end{document}